\documentstyle[12pt,aps]{revtex}

\begin{document}

\title{Phase boundary dynamics in a 
one-dimensional non-equilibrium lattice gas
}

\vspace{0.3in}

\author{ G.Oshanin$^1$, J.De Coninck$^2$,
M.Moreau$^1$ and S.F.Burlatsky$^{3\dag}$
}

\vspace{0.3in}

\address{$^1$  
 Laboratoire de Physique Th\'eorique des Liquides,
 Universit\'e Pierre et Marie Curie,
4, Place Jussieu, 75252 Paris, France}

\vspace{0.2in}

\address{$^2$ Centre de Recherche en Mod\'elisation Mol\'eculaire,
Universit\'e de Mons-Hainaut,
 20, Place
du Parc, 7000 Mons, Belgium}

\vspace{0.2in}

\address{$^3$ Department of Chemistry, 
Massachusetts Institute of Technology,
Cambridge, MA 02139 USA}

\vspace{0.4in}

\bigskip

\begin{tightenlines}

\address{\rm }
\address{\mbox{ }}
\address{\parbox{16cm}{\rm \mbox{ }\mbox{ }
We study  dynamics of a phase boundary in 
a one-dimensional lattice gas,  which is 
initially put into a non-equilibrium  
configuration and then evolves 
in time by particles 
performing 
 nearest-neighbor  random 
walks constrained by hard-core interactions. Initial non-equilibrium
configuration is characterized by an 
$S$-shape density profile,
such
that particles density from one side of the origin (sites $X \leq 0$)
is larger  (high density phase, HDP) than that from the other side (low-density
phase, LDP).   
We suppose that all lattice gas particles, except 
for the rightmost particle of the HDP, 
have symmetric hopping
probabilities.  The 
rightmost particle of the HDP,  which determines the position of 
the phase separating
boundary, is subject to a constant force $F$, oriented
towards the HDP; in our model this force mimics an effective 
tension of the phase separating
boundary.  We find that, in the general case, 
the mean displacement $\overline{X(t)}$
of the phase
boundary grows with time as $\overline{X(t)} = \alpha(F) t^{1/2}$, where 
the prefactor $\alpha(F)$
depends on $F$  and on the initial densities in the HDP and LDP.
We show that $\alpha(F)$ can be positive or negative, 
which means that depending on the physical conditions
the HDP may expand or get compressed. 
In the particular  case when $\alpha(F) = 0$, i.e. when 
the HDP and LDP coexist with
each other, 
the second moment of the phase boundary displacement
 is shown to grow with time 
sublinearly, $\overline{X^2(t)} = \gamma t^{1/2}$, where the prefactor $\gamma$
is also calculated explicitly.
Our analytical predictions are shown to be in a very good agreement
 with the results of Monte
Carlo simulations.
}}
\address{\mbox{ }}
\address{\parbox{16cm}{\rm }}
\maketitle

\vspace{0.3cm}

\makeatletter
\global\@specialpagefalse

\makeatother

\hrule
\bigskip
\noindent $^{\dag}$ Present address: 
LSR Technologies, Inc., 898 Main St, Acton, MA 01720-5808
 USA

\end{tightenlines}

\section{Introduction.}

A fundamental question concerning the behavior
 of systems out of
equilibrium can be formulated as follows. Suppose that two different
 phases, composed of the same
or of two different substances, are initially prepared 
in different regions of space 
and have a common interface.  
What is the future evolution for the system and for
the phase-separating interface? This problem appears in
the analysis of 
  such diverse phenomena as 
expansion of the poisoned state in catalytic reactions, or, 
more generally, propagation of chemical
fronts, dielectric breakdown, 
growth of dendrites and clusters in Ising magnets, 
spatial intermittency in hydrodynamics, 
wetting, rise of liquids in capillaries and many others (see
\cite{cross,amar,pomeau,cazabat,lan,coll,collet,spohn,kardar,kapral,arm,ebert}
 and references therein).

Theoretical analysis of the problem follows basically two distinct avenues.
 One type of approach is to describe
the system evolution in terms of some
 appropriate set of starting equations, 
 a standard list of which
includes such non-linear differential field equations of
 different complexity as, e.g. Newell-Whitehead,
viscous Burgers, Swift-Hohenberg, Cahn-Hilliard equations 
and etc
\cite{cross,amar,pomeau,cazabat,lan,coll,collet,spohn,kardar,kapral,arm,ebert}.
 These equations are usually referred
to as "microscopic", with the understanding, however, 
that they don't involve
 atomic degrees of freedom but
rather serve as elementary building blocks
 from which the analysis starts. 
Another approach consists
in the direct study  of  models involving 
particles with microscopically defined  dynamics  \cite{leb,alex,hardy,frisch}. 
The dynamic rules can be, for instance, 
 chosen to construct a suitable
cellular automaton, which converges asymptotically
to the field equation in question, 
 e.g. the Navier-Stokes equation \cite{hardy,frisch,zal,roth},
allowing then
 for much more efficient numerical analysis than simulations of 
the continuous-space counterpart. Alternatively, 
they can be deduced from realistic microscopic interactions
 with the intention to derive 
equations describing the time evolution of some
 macroscopic properties, such as, e.g. local particle densities 
\cite{leb,gia,giac,guo}. 
Considerable progress has been made recently in 
this direction \cite{leb,gia,giac,guo}, 
which revealed, however, the fact that macroscopic
 equations derived on the basis of
realistic microscopic interactions may have a 
different structure compared to the generally accepted
field equations and can be 
reduced to them only under certain assumptions.

In the present paper\footnote{This paper is based partly 
on the talk given at the conference
on Inhomogeneous Random Systems, Palaiseau, France, January 1997, and
at the workshop on Instabilities and Non-Equilibrium
Structures, Santiago, Chili, December 1997}   
we study dynamics of
the phase boundary 
in a  two-phase microscopic 
model system  
consisting of identical hard-core
particles which are placed
initially on
 a one-dimensional, infinite
in both directions lattice  
 in
 a non-equilibrium, "shock"-like configuration.
That is, particles mean densities from the left
and from the right of the origin
 of the lattice (Fig.1), which we denote as
$\rho_{-}$ and $\rho_{+}$ respectively,
are generally not equal to each other. We suppose, without lack of generality that $\rho_- \geq
\rho_+ \geq 0$, and will call in what follows the phase which initially occupied the
left half-line as the high-density phase (HDP), while the phase initially occupying
the right half-line will be referred to as the low-density phase or the LDP. 
Particles are then allowed to perform
$symmetric$, (i.e. with equal probabilities for going to the left or to the right),
hopping motion
between the 
nearest lattice sites under the
constraint that neither two
particles can simultaneously
occupy the
same lattice site and can not
pass through each other. 
Further on, we single out the rightmost particle 
of the HDP, 
which determines 
position of
the phase boundary and which 
we will call in the following
as the PBP - 
the "phase boundary" particle (Fig.1). We suppose that 
this only particle
is subject to
a constant force $F$ which favors the PBP to jump in a preferential direction. 
Thus for the PBP the probabilities of
going to the right ($p$) and to the left ($q$) will be different 
from each other.  In most situations we will suppose that $F$ is directed towards
the HDP; we adopt the convention that in this case
$F$ is positive definite, $F \geq 0$; the PBP hopping probabilities
$p$ and $q$ 
are related to the force 
and the reciprocal temperature $\beta$
 through $p/q = exp( - \beta F)$ and
$p + q = 1$.
 From the physical point
of view, such a $constant$ force can be understood
as an effective boundary tension  derived from the
solid-on-solid-model Hamiltonian of the
 phase-separating boundary 
 \cite{joela,joelb,joelc} and 
mimic, in a mean-field fashion, the presence of
attractive interactions
between the lattice-gas particles  
which are not explicitly included into the model. 

We hasten to remark 
that the system evolution 
 in the case when
long-range attractive interactions
between the gas particles are present can
 be fairly more complex. First, in this case the
hopping probabilities of any 
given particle are coupled
to the instantaneous positions of 
all other gas particles and, consequently,
evolution of the local densities is described by  
non-linear and non-local integro-differential equations
(see  \cite{gia,giac} and references therein). 
Furthermore,  
  the "boundary-tension" force $F$ is generally time-dependent and
reaches a constant value only at sufficiently large 
times. Moreover,  this value  will be explicitly dependent   on the
density distribution around the PBP resulting
in a non-linear coupling 
between the dynamics of the PBP
and of the lattice-gas particles.
A mean-field-type analysis  of this situation, 
which has interesting applications
within the context 
of spreading of liquid monolayers
on solid supports \cite{burc,burd}, has been presented recently 
in \cite{force}; here, we 
assume  that $F$ is a
fixed given
 parameter, which is independent
of the particle density. We note that such an assumption is justified 
when particle-particle interactions are sufficiently weak and time $t$ is sufficiently
large \cite{force}.

We note now that several particular cases
of the general model under study have been already
discussed in the literature. Consequently, 
 considering different limiting
with respect to 
$\rho_{+}$, $\rho_{-}$ and $F$ situations
 we will be able to check our
predictions against already known results. 
We mention 
some of these results:

(a)  Dynamics of the PBP in the symmetric 
case $\rho_{-} =  \rho_{+} = \rho$ and
in the absence
of an external force $F$ 
has been studied as early as
1965 by Harris  \cite{har}, who has shown rigorously that the
mean-square displacement of the PBP grows sublinearly with time,
\[\overline{X^{2}_{r}(t)} \; = 
\; \frac{1 - \rho}{\rho} \; \sqrt{\frac{2 t}{\pi}}   \quad (1)\]
Here and henceforth
 the overline denotes the averaging with respect to different realizations of
the PBP trajectories $X_{r}(t)$. 
Moreover, it was shown in \cite{ara} that $t^{-1/4} X_{r}(t)$ converges 
in distribution to a
Gaussian variable with variance $(1 - \rho) \sqrt{2/\pi}/\rho$.

(b) Further on,  a rigorous probabilistic
description of the situation with
  $\rho_{-} = 1$, $\rho_{+} = 0$  and zero "boundary tension"
 force has been developed in \cite{ara}. 
It was proven
that in this case
the mean displacement of the PBP
 grows in time in proportion to $\sqrt{t \; log(t)}$ as $t \to \infty$. 

(c) An opposite case when $F = \infty$, (such that the PBP performs totally
directed random walk), while 
$\rho_{+} = 0$ and $\rho_{-} = \rho$
has been considered in 
\cite{bura}. It was demonstrated that the mean displacement of the PBP obeys
$\overline{X_{r}(t)}  =   - \alpha_{lim} \sqrt{t}$,
in which law the prefactor
 $\alpha_{lim}$ is defined implicitly by
\[\sqrt{\frac{\pi}{2}}  \; \alpha_{lim} \; \exp(\alpha^{2}_{lim}/2) \; [1 \; - \;
\Phi(\alpha_{lim}/\sqrt{2})] 
\; = \; 1 - \rho, \quad (2)\]
where $\Phi(x)$ denotes the error function. 

(d)  A more general situation has been considered
in
\cite{burb} and subsequently, in \cite{olla}, which works 
deal with the behavior of the driven PBP
in the symmetric case $\rho_{-} = \rho_{+} = \rho$.
 Both works have shown that at arbitrary negative values of the
boundary tension force, $F \leq 0$, the mean-square
displacement of the SFB follows 
\[\overline{X_{r}(t)}  \; =  \; \alpha(F) \; \sqrt{t}   \quad (3)\]
In \cite{burb} it was found analytically, in terms of a mean-field-type approach,
and also confirmed by numerical Monte Carlo simulations  
that for arbitrary $\rho$ and $p \geq q$ the parameter
 $\alpha(F)$ is determined by the following
transcendental equation:
\[  ( \sqrt{\frac{\pi}{2}} \; \alpha(F) \; \exp(\alpha^{2}(F)/2) \; [1 \; + \; 
\Phi(\alpha(F)/\sqrt{2})] \; + \quad\]
\[+  \; \frac{p - q (1 - \rho)}{p - q}) 
( \sqrt{\frac{\pi}{2}} \; \alpha(F) \; \exp(\alpha^{2}(F)/2) \; \times \quad\]
\[ \times  [1 \; - \; 
\Phi(\alpha(F)/\sqrt{2})] \; + \; \frac{q - p (1 - \rho)}{p - q}) \; = \; 
\frac{p q \rho^{2}}{(p - q)^{2}}  \quad (4)\]
In \cite{olla}, which has also established an interesting 
 relation between the time evolution of
a symmetric  lattice gas with a single driven tracer and the evolution
 of the interfaces in a two-dimensional Potts
model with Glauber dynamics,  the PBP dynamics was analysed 
in terms of a rigorous probabilistic
approach and 
the result in Eq.(4) has been rigorously proven.

An interesting observation made in \cite{burb} 
and subsequently, in \cite{olla}, 
concerned the validity of the
Einstein relation for the tracer diffusion 
in a one-dimensional 
hard-core lattice gas.
An illuminating discussion of this issue and 
a considerable amount of new results establishing the Einstein
relation 
for different interacting particle systems
 was presented recently in \cite{fer}.
Now, Eq.(4) shows 
that in the case of a weak asymmetry, i.e. when $p - q \to 0$,
the parameter $\alpha(F)$ is given exactly by
\[\alpha(F) \; = \; (p \; - \; q) 
\; \frac{1 \; - \; \rho}{\rho} \; \sqrt{\frac{2}{\pi}}  \quad (5)\]
If one defines then the time-dependent mobility $\mu(t)$ of the driven PBP as
$\mu(t) = lim_{F \to 0} X(t)/ F t$, it would
yield, by virtue of Eq.(3), that
$\mu(t) = t^{-1/2} lim_{F \to 0} \alpha(F)/F$, 
where $\alpha(F)$ is given by Eq.(5). On the other hand, the time-dependent
diffusivity $D(t)$ of the PBP in absence of external force obeys $D(t) 
= (1 - \rho)/\rho ( 2 \pi t)^{1/2}$ \cite{har}. As it was shown in \cite{burb}
and \cite{olla}, the asymptotic form in Eq.(5) implies that $\mu(t) = \beta D(t)$,
i.e. the Einstein relation holds exactly 
in the non-stationary regime\footnote{Two of us, G.O. and S.F.B., wish to thank
Professor J.L.Lebowitz who has suggested us to examine the question of validity of the
Einstein relation in the non-stationary regime.}.  

\vspace{0.5cm}    

Here we  
develop an analytical dynamical description of the lattice gas model with initial
"shock"-like configuration of particles, which is 
based on the mean-field-type assumption 
that the average of the product 
of local realization-dependent variables, describing occupation of lattice
sites, factorizes into the product of their average values, i.e. the local densities. 
Such an assumption permits us to derive the 
closed-form system of equations for the time evolution of the PBP mean displacement
and of the density profiles around it. These equations 
allow for the analytical solution and explicit computation of the time evolution
of the PBP mean displacement. 
Our main results are the following:

 We find that at sufficiently large times 
the mean displacement of the PBP obeys 
 Eq.(3), in which the prefactor $\alpha(F)$ is determined
implicitly as the solution of the transcedental equation
\[q \; \rho_{-} \; \{1 \; + \; \sqrt{\pi/2} \; \alpha(F) \; \exp(\alpha^{2}(F)/2) \; 
 [1 \; + \; sgn(\alpha(F)) \; \Phi(|\alpha(F)|/\sqrt{2})]\}^{-1} \; - \; \quad\]
\[- \; p \; \rho_{+} \; \{1 \; - \; \sqrt{\pi/2} \; \alpha(F) \; \exp(\alpha^{2}(F)/2) \;
 [1 \; - \; sgn(\alpha(F)) \; \Phi(|\alpha(F)|/\sqrt{2})]\}^{-1}  \; = \; q \; - \; p,    \quad (6)\]
where $sgn(\alpha) = 1$ for $\alpha > 0$ and $sgn(\alpha) = - 1$ for $\alpha \leq 0$. Eq.(6)
holds for any relation between $p$, $q$ 
and $\rho_{\pm}$ and reduces to the previously
obtained
Eq.(2) and Eq.(4) in the  
appropriate limits. This result has been found subsequently in \cite{olla}.

 Equation (6) predicts that three
different regimes can take place depending on the relation
between $p/q$ and $\rho_{\pm}$:

(1)  When $p (1 - \rho_{+}) > 
q ( 1 - \rho_{-})$ the parameter $\alpha(F)$ is finite and
positive definite, which means that the
HDP expands compressing the LDP.
In the particular case $p/q = 1$ and $\rho_{+} = 0$ the parameter
$\alpha(F)$ appears to be a positive, logarithmically growing
with time function, which behavior  agrees
 with the results of \cite{ara}.

(2) When $p (1 - \rho_{+}) = 
q ( 1 - \rho_{-})$, the parameter $\alpha(F) = 0$.
This relation between the system parameters when 
the HDP and the LDP are in equilibrium with
each other and the PBP mean displacement is zero, was found 
also in \cite{fer} from the analysis of the stationary behavior
in a finite one-dimensional lattice gas  (see also Section VI.A). 
Despite the fact that the PBP mean displacement is zero, the
fluctuations in the PBP position grow with time.
We show here that in this case the mean-square
displacement of the PBP obeys
\[\overline{X_{r}^{2}(t)} \; = 
\; \frac{(1 - \rho_{-})(1 - \rho_{+})}{(\rho_{-} 
\; + \; \rho_{+} \; - \; 2 \; \rho_{-} \; \rho_{+})} 
\; \sqrt{\frac{8 t}{\pi}},   \quad (7)\]
which reduces to the classical result in Eq.(1) in the limit
$p = q$ and $\rho_{-} = \rho_{+}$. Eq.(7) 
is derived here using heuristic arguments based on the Einstein
relation between the mobility and diffusivity of
 a test particle and is confirmed by Monte Carlo
simulations.

(3) When 
$p (1 - \rho_{+}) <  
q ( 1 - \rho_{-})$ the parameter $\alpha(F)$ is less than zero
- the expanding LDP and the applied force effectively
compress the HDP. 

Further on, we
show that 
the particles density profile
as seen from the moving PBP, stabilizes
around its position, approaching a constant value. 
We  find
that in the regime when the HDP expands (i.e. $\alpha(F) > 0$)
the
density profile $\rho(X;t)$ around the PBP is given by
\[\rho(X < X(t);t) \; \approx \; \frac{\rho_{-}}{1 +
I_{+}(|\alpha(F)|)}  \; [1 \; + \; \alpha(F) \; t^{-1/2} (X(t) - X) \; + \; ... \;
], \quad (8.a)\]
\[\rho(X > X(t);t) \; \approx \; \frac{\rho_{+}}{1 -
I_{-}(|\alpha(F)|)} \; 
[1 \; + \; \alpha(F) \; t^{-1/2} (X(t) - X) \; + \; ... \;
], \quad (8.b)\]
for $X(t) - X \ll \sqrt{t}/A$. Within the opposite limit ($\alpha(F) < 0$) 
when the
HDP gets compressed,  $\rho(X;t)$ follows
\[\rho(X < X(t);t) \; \approx \; \frac{\rho_{-}}{1 -
I_{-}(|\alpha(F)|)} \; [1 \; + \;\alpha(F) \; t^{-1/2} (X(t) - X) \; + \; ... \;
], \quad (9.a)\]
\[\rho(X > X(t);t) \; \approx \; \frac{\rho_{+}}{1 +
I_{+}(|\alpha(F)|)} \; [1 \; + \; \alpha(F) \; t^{-1/2} (X(t) - X) \; + \; ... \;
], \quad (9.b)\]
where
\[I_{\pm}(|\alpha(F)|) \; = \; \alpha^{2}(F) \; \int^{\infty}_{0} dz \; \exp(-
\frac{\alpha^{2}(F)}{2} \; (z^{2} \mp 2 z) \; = \quad\]
\[ = \; \sqrt{\frac{\pi}{2}} \; |\alpha(F)| \;
 \exp(\frac{\alpha^{2}(F)}{2}) \; 
[1 \; \pm \; \Phi(|\alpha(F)|/\sqrt{2})]  \quad (10)\]

This paper is outlined as follows: in Section 2 we formulate our model
and discuss the approximations involved. In Section 3 
we write down basic equations, describing the time evolution of the
PBP and of the lattice-gas particles 
in the particular case
when the LDP is absent. 
Section 4 presents
solutions of the dynamic equations in the case $\rho_{+} = 0$ 
and discussion of the PBP dynamics at
different values of  
the initial mean density $\rho_{-}$
and  of the "boundary tension" force $F$. 
Further on, in
Section 5  we consider the general case when the LDP is present
and 
evaluate the dynamical equations describing the time behavior of the system under study.
Section 6 is devoted to the analysis of these equations and evaluation of their
solutions.
Next, Section 7 
contains
the description of the Monte Carlo 
simulations algorithm. Finally, 
in Section 8 we conclude with a brief
summary of results and discussion.

\section{The model.}

Consider a one-dimensional, infinite in both directions
regular lattice of unit spacing, 
the sites $\{X\}$ of which are partly occupied by identical
particles.
Suppose next that the initial configuration of 
particles is as 
depicted in Fig.1, i.e.
all particles are placed on the lattice with 
the single occupancy condition, at random positions 
and in such a way that
the mean particle  density ($\rho_{-}$)
at sites $X \leq 0$ is different from the 
mean particle density ($\rho_{+}$) at the sites with $X > 0$.
As we have already mentioned, we supposed that
$\rho_{-} \geq \rho_{+} \geq 0$ and call the gas phase from the left of the PBP
as the HDP and the gas phase from the right of the PBP as the
LDP.
 
After deposition onto the lattice, the particles are allowed to move
by attempting jumps to neighboring sites. The motion is 
constrained by the
hard-core exclusion between the particles; that is, neither two particles
can simultaneously occupy the same lattice site nor can pass through each other.
For a given realization of the process 
the instantaneous particle configuration is 
described by
an infinite set of time-dependent
occupation variables $\{\tau_{X}(t)\}$,  
where each $\tau_{X}(t)$  assumes
 two 
possible values; namely, $\tau_{X}(t) = 1$
if the site $X$ is occupied at time  $t$
 and $\tau_{X}(t) = 0$ if this site is vacant. 
Position  of the PBP
at time  $t$ is denoted as
 $X_{r}(t)$, which is also a random, 
realization-dependent function.

More specifically, we define particles
dynamics as follows: each particle waits a random, 
exponentially distributed time with mean
$1$ and then selects, with given probabilities,
the  direction of jump - to the right or to the left. 
When the direction of 
the jump is chosen, the
particle attempts to jump onto  the nearest site. 
If the target site is unoccupied by
any other particle at this moment of time, the jump
 is instantaneously fulfilled. 
If the site is occupied, the particle remains
at its position and waits till the next attempt.
The process is memory-less and the choice of jump directions
for different attempts is uncorrelated.

We will distinguish between
the jump probabilities of the PBP
and the jump probabilities of all other particles of the gas.
The jump probabilities of the gas
 particles are $symmetric$, i.e. 
for them an attempt to jump to the right and an attempt to jump  
to the left occur with probability 
 $1/2$, while the jump probabilities of the 
PBP are $asymmetric$;
it attempts to jump to the right 
with probability $p$ and to the left
with probability $q$, $p + q = 1$.  These probabilities
are related to the "boundary tension" force and the
temperature $T = 1/\beta$ through the relation 
$p/q = exp( - \beta F)$.

Here we will develop a mean-field-type description of
the time evolution of the system under study
using an approximate picture, based on two
assumptions: 

We assume first that the average of the product of the occupation variables
$\tau_{X}(t)$ of different sites
factorizes into the product of their average values,
which corresponds to the local equilibrium assumption\footnote{Note, that the density
profiles around the PBP, as shown by Eqs.(8) and (9), tend to constant values on
progressively larger and larger scales as the PBP advances; that is, there is no
structure in the density 
profiles and they are merely the product measures. This "propagation of local
equilibrium" \cite{spohn2} insures that the decoupling procedure involved is correct
in a large time scale.}.
Under such an assumption we can describe the system evolution
directly in terms of the realization-averaged values
$\rho(X;t) = <\tau_{X}(t)>$, which define the local density
of the site $X$ at time $t$ or, in other words, the
probability 
that the site $X$
is occupied by a lattice gas particle at time  $t$.
The resulting equations will then be  
closed with respect to $\rho(X;t)$, i.e. will not 
include  higher-order correlation function. A
non-trivial aspect of these equations, which is common for diverse
 front propagation problems
\cite{coll}, is that  one of the boundary conditions is imposed 
in the moving frame.

Secondly, the evolution of the
spatial position of the phase boundary will
be  described in terms of $P(X;t)$, which
defines the probability of having the PBP 
at site $X$ at time  $t$. Anticipating that in the limit $t \to \infty$ the ratio
\[\frac{\overline{X_{r}(t)}}{\sqrt{\overline{X_{r}^{2}(t)}}}
 \; \to \; \infty,  \quad (11)\]
i.e. that fluctuations in the PBP trajectory grow at essentially slower rate
than its mean displacement, we will neglect fluctuations in
the PBP trajectories, supposing 
 that the 
position  of 
the PBP at time $t$ is a well-defined function of time, 
which is the same for all realizations of the process, $X_{r}(t) = X(t)$. We note that
such an assumption is quite consistent with rigorous 
results presented in \cite{olla}. 

These two simplifying assumptions will allow us to determine explicitly the dynamics
of the PBP  and to calculate 
 the density profiles as seen from the PBP. 
Our analytical predictions 
 will be checked  against the results of Monte Carlo
simulations of the stochastic process, described in the beginning of this Section,
which allows a direct control of the validity of 
the local equilibrium assumption. 

\section{Dynamical equations in the absence of the low-density phase.}

We consider first dynamics of the PBP in the particular case when the LDP is absent
(Fig.2),
i.e. $\rho_{+} = 0$, and thus the jumps of the PBP away from the HDP are
unconstrained. 

We start with the derivation of the dynamical equations, 
which govern the time evolution of
$\rho(X;t)$. 
In the continuous-time limit
and under the assumption of the factorization of the occupation variables 
 at different sites, dynamics of $\rho(X;t)$ is
guided
by the following balance equation
\[\dot{\rho}(X;t) \; = \; - \; \frac{1}{2} \; \rho(X;t) \;  \{1 \; - \; \rho(X+1;t) \; 
+ \; 1 \; - \; \rho(X-1;t)\}  \; + \quad\]
\[+ \; \frac{1}{2} \; (1 \; - \;  \rho(X;t) ) \; \{\rho(X+1;t) \;  +
\; \rho(X-1;t)\}, \quad (12)\]
where the terms in the first two lines describe the contribution due to the jumps from the
 occupied site $X$
onto the neighboring unoccupied sites, 
while the terms in the third line account for
possible arrivals of particles 
to unoccupied site $X$ from the occupied adjacent sites.  
One may readily notice that in Eq.(12) non-linear terms cancel each other and it reduces
to the discrete-space diffusion equation
\[\dot{\rho}(X;t) \; = \; \frac{1}{2} \;  \{\rho(X+1;t) \; + \;  \rho(X-1;t)
 - \; 2 \; \rho(X;t)\} \quad (13)\]

We note now that in  deriving Eq.(13) 
we have implicitly supposed that the jump probabilities of
particles arriving from the site $X + 1$ are  symmetric.
This means that Eq.(13)  holds only for the sites $X$, which are inaccessible
 for the PBP, whose jumping probabilities are asymmetric by definition. 
Consequently, Eq.(13) is valid only  for the sites $X$ 
such that $X <  X(t) - 1$. For $\rho(X(t) - 1;t)$
 we have instead of Eq.(12)
\[\dot{\rho}(X(t) - 1;t) \; = \; \frac{1}{2} \;  \{ \rho(X(t) - 2;t) \; (1 \; - \;  \rho(X(t) - 1;t))
\; - \;  \quad\]
\[- \; \rho(X(t) - 1;t)  \; (1 \; - \; \rho(X(t) - 2;t)) \} + \; \quad\]
\[  + \; q \; (1 \; - \; \rho(X(t) - 1;t)) \;  \rho(X(t)-2;t) \;  - 
\;  p \; \rho(X(t) - 1;t) \;  \quad (14)\]
The terms in the first two lines of Eq.(14) describe  exchanges
 of particles between the sites $X(t)-1$
and $X(t) - 2$. The particles which may be involved in 
these exchanges are the lattice gas
particles which have symmetric jumping probabilities and
 here, again, the non-linear terms
cancel each other. 
Two terms in the third line of Eq.(14) account for the effective change
in the occupation of the $(X(t) - 1)$-site due to jumps of the PBP and are defined
in the frame of reference moving with the phase boundary. The first
term describes creation of a vacancy at a previously occupied site $(X(t) - 1)$ 
due to the unconstrained
jump of the PBP away of the gas phase. The second one accounts for 
the effective creation
of a particle at $(X(t) - 1)$ in the event 
when the PBP jumps onto unoccupied site $(X(t) - 1)$ 
and the site $(X(t) - 2)$ is occupied
prior to the jump. 

Similar reasonings yield the following equation for the
 time evolution of the probability
distribution $P(X;t)$, which defines the PBP dynamics,
\[\dot{P}(X;t) \; = \; -  \; P(X;t) \; \{ \; p  \; + \; 
q \; ( 1 \; - \; \rho(X(t) - 1;t))\} \; + \; \quad\]
\[ + \;  p  \; P(X-1;t) \; +  \; q \; ( 1 \; - \; \rho(X(t);t)) \; P(X + 1;t) \quad (15)\]
Multiplying both sides of Eq.(15) by $X$ and summing over all lattice sites we find that the
displacement of
the PBP obeys
\[\dot{X}(t) \; = \; p \; - \; q \; + \; q \; f(1;t), \quad (16)\]
where we took into account
the normalization condition $\sum_{X} P(X;t) = 1$ and 
denoted as $f(\lambda;t)$
the pair-wise correlation function
\[f(\lambda;t) \; = \; \sum_{X} P(X;t) \; \rho(X - \lambda;t) \quad (17)\]
Equation (17) defines  the probability of having at time moment $t$
a particle at distance $\lambda$ from the PBP, or, in other words, 
can be interpreted as the density profile as
seen from the moving PBP.

We now turn  to the time evolution of $f(\lambda;t)$. 
Differentiating the pair-wise correlation function in Eq.(17)
with respect to time, we have
 \[\dot{f}(\lambda;t) \; = \; \sum_{X} \{ \dot{\rho}(X - \lambda;t)  
\;  P(X;t) \; + \; \dot{ P}(X;t)  \;  \rho(X - \lambda;t)\} \; \quad (18)\]
We notice now 
that again the behavior for $\lambda = 1$ and $\lambda > 1$ has to
be  considered  separately.
In the domain $\lambda > 1$ we find, taking advantage of Eqs.(13) and (15), 
that $f(\lambda;t)$ obeys
 \[\dot{f}(\lambda;t) \; = \; \frac{1}{2} \; \{f(\lambda - 1;t) \; + 
\; f(\lambda + 1;t)
\; - \; 2 \; f(\lambda;t)\} \; - \; \quad\]
\[ - \; (p - q) \; f(\lambda;t) \; + \; p \;  f(\lambda-1;t) \; + \; q 
\; f(\lambda+1;t) \; - \quad\]
\[ - \; q \; \sum_{X} P(X;t) \; \rho(X-1;t) \; \rho(X - \lambda - 1;t)  \; + \quad\]
\[+ \; q \; \sum_{X} P(X;t) \; \rho(X-1;t) \; \rho(X-\lambda;t)  \quad (19)\]
We proceed further on making the same simplifying 
assumption, which underlies the derivation of Eqs.(12) to (15), i.e. assuming
that 
the average of the product of 
the occupation variables  decouples into the product of the average values.
This means that the two 
last terms in Eq.(19) can be rewritten as
\[  \sum_{X} P(X;t) \; \rho(X-1;t) \; \rho(X - \lambda - 1;t) \; 
= \; \quad\]
\[=  \{\sum_{X} P(X;t) \;
\rho(X-1;t)  \} \;   \{\sum_{X'} P(X';t) \; \rho(X'-\lambda-1;t)  \}, 
 \quad (20.a)\]
and 
\[  \sum_{X} P(X;t) \; \rho(X-1;t) \; \rho(X - \lambda ;t) \; 
= \; \quad\]
\[= \; \{\sum_{X} P(X;t) \;
\rho(X-1;t)  \} \;  \{\sum_{X'} P(X';t) \; \rho(X'-\lambda;t) \} 
 \quad (20.b)\]
Decoupling of the third-order correlation functions as in Eqs.(20) permits 
us to cast Eq.(19)  into the following form
 \[  \dot{f}(\lambda;t) \; = \; \frac{1}{2} \; 
 \{ f(\lambda - 1;t) \; + \; f(\lambda + 1;t)
\; - \; 2 \; f(\lambda;t) \} \; - \; \quad\]
\[- \;  f(\lambda;t) \; + \; p \; f(\lambda-1;t) \; + \; q \; f(\lambda+1;t) \; - \quad\]
\[ - \; q \; f(1;t) \; ( f(\lambda+1;t) \; - \;  f(\lambda;t)) \quad (21)\]
Equation (21) does not include now the third-order correlation
 functions and thus is closed with respect to
$f(\lambda;t)$. 

Next, using Eqs.(18), (15) and (14) and decomposing the
 third-order correlation
functions into the product of pair-wise correlations, we obtain for 
the time evolution of $f(1;t)$: 
\[ \dot{f}(1;t) \; = \; \frac{1}{2} \;  \{f(2;t) \; - \; f(1;t)  \} \; - \quad\]
\[\; - \; f(1;t) \; + \; 2 \; q \; f(2;t) \; (1 \; - \; f(1;t)) \; + \quad\]
\[+ \; p \; (f(0;t) \; - \; f(1;t)) \; + \; q \; f^{2}(1;t) \quad (22)\]
From Eq.(22) we can now deduce 
 the boundary condition for Eq.(21). Setting in Eq.(21) the correlation 
parameter $\lambda$ equal to $1$ and comparing the terms in the rhs of Eq.(21)
against
the terms in the rhs of Eq.(12) we can infer that $f(1;t)$ obeys
\[\frac{1}{2} \; (f(1;t) \; - \; f(0;t)) \; = \; p \; f(1;t) \; -  \; q \;  f(2;t) 
\; (1 \; - \; f(1;t))   \quad (23.a) \]

Another pair of boundary conditions will be 
\[\left . f(\lambda;0) \; = \; (\sum_{X} P(X;t) \; \rho(X - \lambda;t))\right|_{t=0}
 \; = \; \rho_{-}, \quad (23.b)\]
and
\[f(\lambda \to \infty;t) \; \to \; \rho_{-}, \quad (23.c)\]
which mean that initially the lattice gas particles are
 uniformly distributed, with mean density $\rho_{-}$,
 on the half-line $X < 0$, and that the density
 of the lattice gas at large separations from
the phase boundary is equal to its unperturbed value.

Equations (16) and (21) to (23) constitute a closed system
 of equations which allows a complete determination
 of
$X(t)$. Solution of these equations will be discussed
 in the next section.

\section{Solution of dynamical equations in the case $\rho_{+} = 0$.}

We now turn  to the continuous-space limit
 and rewrite our equations
 expanding $f(\lambda \pm 1;t)$
into the Taylor series 
and retaining terms up to the second order in powers of the lattice spacing.
We then find  that $f(\lambda;t)$ obeys
\[\dot{f}(\lambda;t) \; = \; \frac{1}{2} 
\; \frac{\partial^2 }{\partial \lambda^2} f(\lambda;t) \;
- \; \dot{X}(t) \; \frac{\partial}{\partial \lambda} f(\lambda;t),   \quad (24)\]
while Eq.(23.a) transforms to
\[\left . \frac{1}{2} \; \frac{\partial}{\partial \lambda} f(\lambda;t) \right|_{\lambda = 1} \; =
\;  \dot{X}(t) \; f(1;t), \quad (25)\]
where, by virtue of Eq.(16),
 we have replaced the multiplier $(p - q + q f(1;t))$ by $\dot{X}(t)$. We note
that Eqs.(24) and (25) hold for any
relation between $p$ and $q$ (any orientation of the force $F$),
but in the absence of the LDP the analysis of the case $p > q$ does not make any sense. 
Consequently, 
in this section  we will consider only the case when $p \leq q$.

\subsection{Expansion of the gas phase.}

Let us first consider the solution of Eqs.(24) and (25) 
supposing  that $X(t) > 0$, ($X(0) = 0$). Conditions when such a behavior takes place will be defined
below.  
We notice that the structure of Eqs.(24) and
(25) calls for the scaling solution in terms of 
variable $\omega = (\lambda - 1)/X(t)$; $0 \leq
\omega \leq \infty$. In terms of
 this variable Eqs.(24) and (25) can be rewritten as
\[\frac{\partial^2 }{\partial \omega^2} f(\omega)  \; + 
\; (\frac{d}{d t} X^2(t)) \; (\omega - 1) \;
\frac{\partial }{\partial \omega} f(\omega) \; = \; 0, \quad (26)\]
and 
\[\left . \frac{\partial }{\partial \omega} f(\omega) \right|_{\omega = 0} \; = \; 
(\frac{d}{d t} X^2(t)) \; f(\omega = 0), \quad (27)\]
while Eqs.(23.b) and (23.c) collapse into a single equation
\[f(\omega = \infty) \; = \; \rho_{-} \quad (28)\]

Solution of Eqs.(26) to (28) can be readily obtained in an explicit form
if we assume that
 $d X^2(t)/d t = A^2$, where $A$ is a time-independent constant, $0 \leq A < \infty$. 
Such an
assumption actually makes sense if we recollect results of \cite{ara} and \cite{bura,burb,olla},
which demonstrated that in two extreme situations, 
i.e. when $p/q = 0$ (totally directed walk of the PBP) and when $p/q = 1$ (no force exerted on the
PBP), the PBP displacement shows the same 
generic behavior $X(t) \sim \sqrt{t}$.
Hence, 
one can expect that for arbitrary $p/q$,
$0 \leq p/q \leq 1$, the PBP displacement should also grow in proportion to
$\sqrt{t}$. 

\vspace{0.5cm}

{\bf Time-independent $A$ ($p/q > 1$).} The general solution of Eq.(26) has the form
\[f(\omega) \; = \; C_{1} \; \int^{\omega}_{0} dz \; \exp( - \frac{A^2}{2} \; (z^2 - 2 z)) 
\; + \; C_{2}, \quad (29)\]
where $ C_{1}$ and $ C_{2}$ are adjustable constants. Substitution of Eq.(29) into  
Eq.(27) gives
\[C_{1} \; = \; A^2 \; C_{2}, \quad (30)\]
while Eq.(28) yields the second relation
\[ C_{1} \; \int^{\infty}_{0} dz \; \exp( - \frac{A^2}{2} \; (z^2 - 2 z)) \; + \; C_{2} \; = \;
\rho_{-} \quad (31)\]
Consequently, we have for the density profile
\[f(\omega) \; = \; \frac{\rho_{-}}{1 \; + \; I_{+}(A)} \;   \{1 \; + 
\; A^2 \;  \int^{\omega}_{0} dz \; \exp( - \frac{A^2}{2} \; (z^2 - 2 z))\}, \quad (32)\]
where the function $I_{+}(A)$ has been made explicit in Eq.(10).

The density $f(\omega)$ in Eq.(32) is a function of $A$, which still remains
undetermined. To define $A$ we notice that $X(t) \sim \sqrt{t}$ behavior and Eq.(16) imply that 
\[f(\omega=0) \; \to \; \frac{q - p}{q}, \; \text{as} \; t \; \to \; \infty, \quad (33)\]
and consequently, we have that in the limit $t \to \infty$ the parameter $A$ approaches a
constant, time-independent value which obeys
\[I_{+}(A) \; = \; \frac{p - q (
1 - \rho_{-} )}{q - p}   \quad (34)\]
Equation (34) implicitly determines $A$ as a function of $p/q$ 
and $\rho_{-}$. Numerical solution of this
equation is presented in Fig.3.

Now, a simple analysis shows that Eq.(34) has
 a unique positive solution for any $p$ and $q$ which
satisfy $p > q (1 - \rho_{-})$, (or, since $p = 1 - q$, such 
$q$ which are less than $1/(2 - \rho_{-})$).
When $p/q \to 1 - \rho_{-}$, the parameter $A \to 0$ as
\[
A \; \approx \; \sqrt{\frac{2}{\pi}} 
\; \frac{p - q (1 - \rho_{-})}{q \rho_{-}}, \quad (35)
\]
and is exactly equal to zero for $p/q = 1 - \rho_{-}$. 
It means that in the domain of parameters such
that $p > q (1 - \rho_{-})$, the gas phase expands and
 the phase boundary moves as $X(t) = A \sqrt{t}$, $A > 0$.
 Before
 we turn to the analysis of the behavior of the PBP in 
the regime $p < q (1 - \rho_{-})$, let us mention some 
other interesting aspects of Eqs.(34) and (32).

\vspace{0.5cm}

{\bf Time-dependent $A$ ($p/q = 1$).} 
We note that Eq.(34) predicts that $A$ diverges 
logarithmically when $p/q \to 1$ (Fig.3). Nameley, 
\[A \; \approx \; \sqrt{- \; 2 \; ln(1 - \frac{p}{q})}, \quad (36)\]
which means apparently that when $p = q$ the parameter $A$ is some 
increasing 
function of time. This is, of course, consistent with the result of \cite{ara} which states
that the mean displacement of the PBP obeys $X(t) \sim \sqrt{t \; ln(t)}$
for the lattice gas with $\rho_{-} = 1$ and $p = q = 1/2$.
Let us now estimate, in terms of our approach,
the  behavior of $A$ for arbitrary $\rho_{-}$ and $p = q = 1/2$. 
For $p = q$ our Eq.(16) reduces to
\[\dot{X}(t) \; = \; f(\omega = 0) \quad (37)\]
Next, supposing that  $X(t)$ still
 follows the law $X(t) = A \sqrt{t}$, in which 
the prefactor  $A$ may be a slowly varying function of time, 
such that $A/\sqrt{t} \to 0$ when $t \to \infty$, we find
that the representation of $f(\lambda;t)$ in terms of a single scaled
variable $\omega$ is still appropriate;  weak time-dependence of
parameter $A$ actually results in the appearence 
of vanishing in time correction terms. We have then
that the boundary condition 
in Eq.(25)  reads
\[\left . \frac{\partial }{\partial \omega} f(\omega) \right|_{\omega = 0} \; \approx \; 
\frac{A^3}{2 \sqrt{t}}, \; \text{when} \; t \; \to \; \infty  \quad (38)\]
On the other hand, we can calculate the derivative of $f(\omega)$ 
directly, using  Eq.(32). This gives
\[\left . \frac{\partial }{\partial \omega} f(\omega) \right|_{\omega = 0} \; \approx \; 
\frac{\rho_{-} A}{\sqrt{2 \pi}} \; \exp( - A^2/2) \quad (39)\]
Comparing next the rhs of Eqs.(38) and (39),  we infer that the parameter $A$ obeys
\[A^2 \; \exp(A^2/2) \; \approx \; \sqrt{\frac{2 \rho_{-}^2 t}{\pi }},  \quad (40)\]
which yields 
\[A \; \approx \; \sqrt{ln(\frac{2 \rho_{-}^2 t}{\pi })} \quad (41)\]
Equation (41) thus generalizes  the result of \cite{ara} for arbitrary 
initial mean density $\rho_{-}$. 

\vspace{0.5cm}

{\bf Wandering of the PBP in the critical case $p/q = 1 - \rho_{-}$.} 
Here we present some
heuristic estimates  of the time evolution of the second moment 
of the distribution $P(X;t)$ in
the case when the gas phase does not "wet" the region $X > 0$, i.e. when $X(t) = 0$. 
To do this,  let us recall 
the Einstein relation between the diffusion 
coefficient $D$ of a particle, which performs an unconstrained symmetric
random walk in absence of external forces, and the mobility $\mu$ of the same particle
in the case when an external 
constant force is present. The Einstein relation states that
$\mu = \beta D$. Of course, it is not clear 
$\it{a priori}$ whether the Einstein relation between
the diffusion coefficient and the mobility should hold also for
 the tracer particle diffusing in a one-dimensional
lattice gas; indeed, it may be invalidated because of the 
hard-core interactions.  
This question has been  addressed for the first time
 in \cite{fer}, in which work several important
 advancements have been made. To illustrate some of the  results obtained in \cite{fer}, 
which are relevant to the model under study, let us first define the mobility of the tracer particle: 
\[\mu \; = \; lim_{t \to \infty} \mu(t),  \quad (42)\]
where $\mu(t)$ denotes
\[\mu(t) \;  =  \; 
 lim_{F \to 0} \frac{X(t)}{F t}, \quad (43)\]
i.e. $\mu(t)$ is the ratio of the mean displacement $X(t)$
of the tracer particle, diffusing in the presence of constant external
force $F$, and $F \; t$; the ratio being 
 taken in the limit when the external force tends
to the critical value (zero) at which the mean displacement vanishes. 
Next, the diffusion coefficient of the tracer particle 
is defined by
\[D \; =  \; lim_{t \to \infty} D(t) \; = \; lim_{t \to \infty} \; 
\{\frac{\overline{X^{2}_{r}(F = 0,t)}}{2 t}\},   \quad (44)\]
where $\overline{X^{2}_{r}(F = 0,t)}$ denotes the mean-square displacement 
of the tracer particle in the case when the external force is
equal to its critical value (zero).
Now, for the 
tracer diffusion in a one-dimensional hard-core lattice gas one has that
$\overline{X^{2}_{r}(F = 0,t)}$ obeys Eq.(1), while $X(t)$ 
is determined by Eqs.(3) and (4).

One readily notices now that the Einstein
relation holds trivially for infinitely large systems, 
 since here both $\mu$ and $D$ are equal to 
zero \cite{fer}.
A more striking result obtained in \cite{fer}  
concerned the case when the one-dimensional
lattice is a closed ring of length $L$. 
It was shown that here both $\mu$ and $D$ are
finite, both vanish with the length of the ring as $1/L$ and 
obey the Einstein relation $\mu(L) =
\beta D(L)$ exactly!  
Next,  \cite{burb} and subsequently, \cite{olla}, 
focused on the non-stationary behavior in infinite systems
and
showed that the Einstein relation
holds in an even more general sense: namely,  
the time-dependent mobility $\mu(t)$ 
and the diffusivity
$D(t)$ obey
\[\mu(t) \; = \; \beta \; D(t), \quad (45)\]
at times $t$ sufficiently large, 
such that the asymptotical
regimes described by  
Eqs.(1) and (3) are 
established.

Now, 
in the situation 
under study we have non-zero critical force ($A = 0$ when $p = q (1 -\rho_{-})$ or, in
other words, when
$F = F_{c} = - \beta^{-1} ln(1 - \rho_{-})$).  We thus define the time-dependent mobility
$\mu(t)$ as 
\[\mu(t) \; = \; lim_{F \to F_{c}} \frac{X(t)}{(F - F_{c}) \; t},  \quad (46)\]
which yields, by virtue of Eq.(35), the following result
\[\mu(t) \; = \; \beta \; \frac{1 - \rho_{-}}{\rho_{-}} \; \sqrt{\frac{2}{\pi t}} \quad (47)\]
Assuming next that the generalized Einstein relation
 in Eq.(45) holds  in the situation under study, 
we find that in the critical 
case  $p = q (1 - \rho_{-})$ the mean-square displacement of
the PBP obeys:
\[\overline{X_{r}^{2}(F = F_{c},t)} \; = \; 
\frac{1 - \rho_{-}}{\rho_{-}} \; \sqrt{\frac{8 t}{\pi}}, \quad (48)\]
which is surprisingly similar 
to the classic result in Eq.(1).

In Fig.4 we compare our analytical prediction 
in Eq.(48) against the results of Monte Carlo simulations,
performed at three different values of the 
gas phase densities $\rho_{-}$.
It shows that our Eq.(48) is in  a good agreement with the
numerical results. This means that
 the Einstein relation in Eq.(45) 
holds even in such a "pathological"
situation, in which the critical value of the external force is not equal to 
zero and
the particle density is different from both sides of the test particle. 

\vspace{0.5cm}

{\bf Density profiles.} Let us now analyse  the form of the density profiles as seen from the
PBP. In Figs.5 and 6 we plot the result in Eq.(32) versus 
the scaled variable $\omega$ for
different initial mean densities and different values of the ratio $p/q$.

 In Fig.5 we depict $f(\omega)$ for fixed
$p/q = 0.9$, which corresponds to fixed "boundary tension" force $F$, and different  initial 
mean densities $\rho_{-}$. In the range of used parameters, all corresponding values of
 $A$ are of the same order ($ A \approx 1$) 
and the density curves look quite similar; starting from the same value at
$\omega = 0$, $f(\omega=0) = 1 - p/q = 0.1$, they quite rapidly, within a few units of
$\omega$,  approach their unperturbed initial values.  We note, however, 
that on the $X$-scale 
it does not mean that the density past the rightmost
 particle rapidly reaches the unperturbed value
$\rho_{-}$.  Instead, at sufficiently large times the density
stays almost constant and equal to $1 - p/q$ 
within a macroscopically large region $\sim X(t)$ .

Now, in Fig.6 we plot $f(\omega)$ versus $\omega$ 
in the opposite case when $\rho_{-}$ is fixed and the "boundary tension"
force is varied. 
Here the density profiles display rather 
strong dependence on the parameter $A$. 
When $A$ is small $f(\omega)$ shows almost linear 
dependence on $\omega$ 
(curves (3) and (4)). The reason for such a behavior is that here 
the phase boundary 
moves essentially slower ($A < 1$), compared to the typical displacements
of the gas particles, which then have
sufficient time 
to equilibrate the density profile past the PBP. 
In the opposite case of 
relatively large values of $A$, ($A > 1$), such 
an equilibration does not take place and the dependence
of $f(\omega)$ on $\omega$ is progressively more pronounced the larger $A$
is.

It may also be  worth-while to discuss the shapes of the density profiles
in terms of the variables $\lambda$ and $t$.  First, from Eq.(32) we have that in the limit
of small $\omega$, i.e. $\lambda \ll X(t)$, the density obeys
 \[f(\lambda;t) \; \approx \; (1 - \frac{p}{q}) \; [ 1 \; + \; \frac{A 
\; (\lambda - 1)}{ \sqrt{t}} \; +
\; ... \; ], \quad (49)\] 
which means that past the PBP the density is almost constant 
in the region whose size
grows in proportion to $X(t)$.
Next, within the opposite limit, i.e. at distances $\lambda$ which
exceed considerably $X(t)$, we obtain from Eq.(32) the following result
\[f(\lambda,t) \; \approx \; \rho_{-} \; - \; (1 - \frac{p}{q}) \; \frac{A \sqrt{2 t}}{\lambda} \; 
\exp(- \frac{\lambda^2}{2 t}) \; + \; ... \;  \quad (50)\]
Equation (50) shows that 
at large separations from the phase boundary  the density approaches 
the unperturbed value $\rho_{-}$
exponentially fast. The approach is
from below  and  is weakly (only through the prefactors) dependent
on the parameters $p/q$ and $A$. 

\vspace{0.5cm}

{\bf Mass of particles and mean density.} 
We close this subsection with a brief analysis  of the time evolution
 of the integral characteristic of the propagating gas phase; namely, 
of the "mass"  $M(t)$ and the mean density $\rho_{mean} = M(t)/X(t)$
of lattice gas particles at sites $X > 0$ at time $t$.

The parameter $M(t)$, which measures the amount of the
gas-phase particles which emerged up to time $t$ in the previously
empty half-line $X > 0$, is formally defined as
\[M(t) \; = \; \int^{X(t)}_{0} dX \; \rho(X;t) \quad (51)\]
Changing the variable of integration, we find that $M(t)$ can be rewritten as
\[M(t) \; = \; \int^{X(t)}_{0} d\lambda \; f(\lambda;t) \; = \quad \]
\[\; = \; X(t) \; \int^{1}_{0} d\omega \; f(\omega) \; = \; M \; t^{1/2}, \quad (52)\]
where $M$ is given by
\[ M \; = \; A \; (1 - \frac{p}{q}) \; exp(A^{2}/2)       \quad (53)\]

Figure 7 displays the plot of the prefactor $M$  versus $p/q$ and shows
 that $M$ is a monotonically
increasing function of $p/q$. In contrast to the parameter $A$, $M$ 
remains finite for $p = q$,
which
means that bulk contribution to the "mass", as it could be expected intuitively, 
comes from the lattice gas particles, whose motion is constrained
 by hard-core exclusions
and whose mean displacement grows only as $\sqrt{t}$, without an additional
logarithmic factor which is specific only to the PBP.

Finally, we depict in Fig.7  the mean density on
 the interval $X \in [0,X(t)]$ , defined as
\[\rho_{mean} \; = \; \frac{M(t)}{X(t)} \; = 
\; (1 - \frac{p}{q}) \; exp(A^{2}/2) \quad (54)\]
Figure 7 shows that despite of the exponential factor $exp(A^{2}/2)$ the
mean density $\rho_{mean}$  rapidly decreases with an
 increase of $p/q$ and is a slowly increasing function of $\rho_{-}$.

\subsection{Compression of the gas phase.}

Let us next address  the question of the PBP dynamics 
in the case
$p < q (1- \rho_{-})$, when $X(t)$ is expected to be less than zero
 and thus the gas phase to be
effectively compressed by the "boundary tension"
 force exerted on the PBP. Recollecting the results of
\cite{bura,burb,olla} we suppose that
 here $X(t)$ obeys $X(t) = - B \sqrt{t}$, $B > 0$, and
define the scaled variable as 
$\omega = (\lambda - 1)/B \sqrt{t}$, where $\omega$ is positive definite
$0 \leq \omega \leq \infty$.
In terms of this variable Eqs.(24) takes the form
\[\frac{\partial^2 }{\partial \omega^2} f(\omega)  \; + \; B^2 \; (\omega + 1) \; 
\frac{\partial }{\partial \omega} f(\omega) \; = \; 0, \quad (55)\]
while the boundary condition in Eq.(25) reads
\[\left . \frac{\partial }{\partial \omega} f(\omega) \right|_{\omega = 0} \; = \; -
B^2 \; f(\omega = 0) \quad (56)\]
Again, the boundary and initial conditions in Eqs.(23.b) and (23.c) 
collapse into a single Eq.(28).

The general solution of Eq.(55) can be written down as
\[f(\omega) \; = \; C_{1} \; \int^{\omega}_{0} dz \; \exp( - \frac{B^2}{2} (z^2 + 2 z)) \; +
C_{2}, \quad (57)\]
where $C_{1}$ and $C_{2}$ are to be chosen in such a way that Eqs.(28) and (56) are satisfied.
Inserting Eq.(55) into Eqs.(28) and (56) we then obtain
\[C_{1} \; = \; - \; B^2 \; C_{2}, \quad (58)\]
and 
\[C_{2} \; = \; \rho_{-} \; \{1 \; - 
\; B^2 \; \int^{\infty}_{0} dz \; \exp( - \frac{B^2}{2} (z^2 + 2 z))\}^{-1} \quad (59)\]
Consequently, the density profile past the PBP 
can be expressed in terms  of $B$ and
$\omega$ as
\[f(\omega) \; = \; \rho_{-} \; \{1 \; - 
\; B^2 \; \int^{\omega}_{0} dz \; \exp( - \frac{B^2}{2} (z^2 + 2 z))\}/ \quad\]
\[/\{1 \; - 
\; B^2 \; \int^{\infty}_{0} dz \; \exp( - \frac{B^2}{2} (z^2 + 2 z))\}  \quad (60)\]
Next, Eq.(16) implies that also in this case $f(\omega=0) \to (q - p)/q$ as $t \to \infty$, which
yields eventually the following closed-form equation for the parameter $B$:
\[1 \; - \; B^2 \; \int^{\infty}_{0} dz \; \exp( - \frac{B^2}{2} (z^2 + 2 z)) \; = \; \frac{q
\rho_{-}}{q - p} \quad (61)\]
Equation (61) can be put into a more compact form if we express
 the integral over $d z$ in terms of  the probability integral.   We then obtain
\[ I_{-}(B)
\; = \; \frac{q (1 - \rho_{-}) - p}{q - p}, \quad (62)\]
where $ I_{-}(B)$ is defined in Eq.(10). In Fig.8 we present
the numerical solution of Eq.(62), plotting the prefactor $B$ 
as a function of the ratio $p/q$ at different
values of the density $\rho_{-}$.

Equation (62) resembles the form of Eq.(34), which determines the parameter $A$, 
but differs from it
in two aspects; first, the rhs of Eq.(62) is exactly the rhs of Eq.(34) but
taken with the opposite sign,
which insures that $B$ is positive for $p < q (1 - \rho_{-})$, 
and second, the sign before the
probability integral in brackets is opposite to that in Eq.(34). 
The latter circumstance
is responsible for the fact that $B$ 
 tends to the limiting value $B_{lim}$ when $q \to 1$ ($p \to 0$).  
When $q \to p/(1-\rho_{-})$ the parameter 
$B$ tends to zero exactly in the same way
 as the parameter $A$ in
Eq.(35) taken with the opposite sign, 
which means that the prefactor in $X(t)$ does not have a
discontinuity at the "critical" point $p/q = 1 - \rho_{-}$
 both for its value and for its slope.  

Now, in the 
limit  $q = 1$ ($p = 0$), Eq.(62) reduces  to
\[\sqrt{\frac{\pi}{2}}  \; B_{lim} 
\; \exp(B_{lim}^2/2) \; [1 \; - \; \Phi(B_{lim}/\sqrt{2})] 
\; = \; 1 - \rho_{-}, \quad (63)\]
which was obtained previously in \cite{bura,burb} (see Eq.(2) in the present paper). 
Within the limit $\rho_{-} \to 1$ Eq.(63) yields
\[B_{lim} \; \approx \; \sqrt{\frac{2}{\pi}} \; (1 - \rho_{-}), \quad (64)\]
which shows that $B_{lim}$, as it could be expected intuitively, 
tends to zero when the density tends to $1$. 

 When the gas is very dilute, i.e. 
$\rho_{-} \ll 1$, we may expect that $B_{lim}$ is large. 
Expanding the probability integral as 
\[ \Phi(B_{lim}/\sqrt{2}) \; \approx \; 1 \; - \; \sqrt{\frac{2}{\pi}} \; B_{lim}^{-1} \;
\exp(-B_{lim}^2/2) \; + \quad\]
\[ + \; \sqrt{\frac{2}{\pi}} \; B_{lim}^{-3} \; \exp(-B_{lim}^2/2) \;  - \; ... \; , \quad (65)\]
we find, upon substitution of Eq.(65) into the Eq.(63), 
the following result
\[ B_{lim} \; \approx \;  \frac{1}{\sqrt{\rho_{-}}},  \quad (66)\]
i.e. $B_{lim}$ diverges when $\rho_{-} \to 0$ in proportion to the inverse of the 
square-root of the particle mean density.  In Fig.9 we present the numerical solution
of Eq.(63)  together with the results of Monte Carlo simulations. Obviously, the agreement is very good.

Finally, in Fig.10 we combine the results of the subsections A and B and plot both analytical
and
 Monte Carlo results obtained 
for the dependence of the prefactor 
$\alpha(F) = X(t)/\sqrt{t}$ on the ratio $p/q$ and the density $\rho_{-}$.
Again, we find very good agreement between our analytical
 predictions and numerical results, which support the validity of the 
approximations involved in our analysis. 

\vspace{0.5cm}

{\bf Density profiles.} Consider now the density profiles as seen from the
 PBP in the compression regime. 
In Fig.11 we plot  $f(\omega)$ versus $\omega$ for 
different values of $p/q$ at fixed $\rho_{-}$.

Figure 11 shows that similarly to the behavior in the expansion regime, 
the density profiles
are quite sensitive to the 
value of the parameter $B$. When $B$ is smaller than unity,
$f(\omega)$ shows almost linear dependence
 on $\omega$, while in the case when $B > 1$ this
dependence is non-linear and $f(\omega)$ 
rapidly drops from $f(\omega=0) = 1 - p/q$ 
to $\rho_{-}$.

We finally present explicit results for $f(\lambda;t)$.
In the limit of small $\lambda$, such that $\lambda \ll X(t)/B^{2}$, we have
\[f(\lambda;t) \; \approx \; (1 - \frac{p}{q}) \; [ 1 \; - \; \frac{B (\lambda - 1)}{\sqrt{t}} \; +
\; ... \; ], \quad (67)\]
which shows that the density is almost constant, (being only slightly  less than $(1 - p/q)$),
in the spatial 
region whose size is of the order of the PBP mean displacement.

For large $\lambda$ we obtain from Eq.(60)
\[f(\lambda;t) \; \approx \; \rho_{-} \; + \; (1 - \frac{p}{q}) \; \frac{B \sqrt{2 t}}{\lambda} \;
exp(- \lambda^{2}/2 t) \; + \; ... \; ,  \quad (68)\]
i.e. similarly to the behavior in the expansion regime, the density approaches the unperturbed value
$\rho_{-}$ exponentially fast and with a
rate which is weakly (only through the pre-exponential factor)
dependent on the parameter $B$ and the ratio
$p/q$.

\section{Dynamical equations in the presence of the low-density phase.}

Let us now consider  the time evolution of the local density $\rho(X;t)$ and of 
the probability
distribution $P(X;t)$ in the general case when the LDP is present and 
$\rho_{-} \geq \rho_{+} \geq 0$.  

One readily notices that also in this case
Eq.(12) and, consequently, Eq.(13), describe the time evolution
 of the realization-averaged
occupation variable 
$\rho(X;t)$ for all $X$ excluding
the sites $X = X(t) \pm 1$. 
Dynamical equation describing evolution 
of $\rho(X;t)$ at $X = X(t) - 1$  will be, however,
 somewhat modified as compared to Eq.(14). 
We have here
\[\dot{\rho}(X(t) - 1;t) \; = \; \frac{1}{2} \; (\rho(X(t) - 2;t) \; 
- \; \rho(X(t) - 1;t)) \; + \quad\]
\[ + \; q \; (1 \; - \; \rho(X(t) - 1;t)) \; \rho(X(t) - 2;t) \; - \quad\]
\[- \; p \; (1 \; - \; \rho(X(t) + 1;t)) \; \rho(X(t) - 1;t),  \quad (69)\]
in which we account that the hops of the PBP
in positive direction can be constrained by the LDP particles
by introducing a factor $(1 \; - \; \rho(X(t) + 1;t))$. In a similar fashion, 
we find that at the site $X = X(t) + 1$ the local particle density obeys
\[\dot{\rho}(X(t) + 1;t) \; = \; \frac{1}{2} \; (\rho(X(t) + 2;t) \; - \; \rho(X(t) + 1;t)) \; - \quad\]
\[ - \; q \; (1 \; - \; \rho(X(t) - 1;t)) \; \rho(X(t) + 1;t) \; + \quad\]
\[+ \; p \; (1 \; - \; \rho(X(t) + 1;t)) \; \rho(X(t) + 2;t)  \quad (70)\]
Next, for the time evolution of the
distribution function $P(X;t)$ we
 obtain the following equation
\[\dot{P}(X;t) \; = \; - \; P(X;t) \; [ p \; (1 \; - \; \rho(X + 1; t)) \; +  \;  q \; (1 \; - \; \rho(X - 1; t))] \; + \quad\]
\[+ \; (1 \; - \; \rho(X; t)) 
\; [  p \; P(X - 1;t) \; + \; q \;  P(X + 1;t) ],  \quad (71)\]
which differs from the corresponding equation of the previous sections, Eq.(15),
by the factors $(1 - \rho(X + 1; t))$ and $(1 - \rho(X; t))$
in the first and third terms respectively;  these factors account, in a mean-field-type fashion,
 for the fact
that hops of the PBP in the positive direction can take place only 
if the corresponding
lattice sites are free of the LDP particles at this moment of time.

Further on, multiplying both sides of Eq.(71) by $X$ and summing over all lattice sites we have that
the mean displacement of the PBP obeys:
\[\dot{X}(t) \; = \; p \; - \; q \; - 
\; p \; f(\lambda = - 1;t) \; + \; q \; f(\lambda = 1;t),   \quad (72) \]
which thus generalizes Eq.(16) for the case of
 non-zero density of the LDP;  the factor $f(-1;t)$ on the right-hand-side of Eq.(72) 
accounts for the hindering effects of the LDP particles on the PBP
dynamics.

Consider now the time evolution of the correlation
 function $f(\lambda;t)$, defined in Eq.(17).
By virtue of Eqs.(18), (13) and (71) we find that the evolution of this property is guided by:
\[\dot{f}(\lambda;t) \; = \; \frac{1}{2} \; [ f(\lambda + 1;t) \; + \; f(\lambda - 1;t) \; -  \;
2 \;  f(\lambda;t) ] \; - \quad\]
\[ - \; f(\lambda;t) \; [1 \; - \; p \; f(-1;t) \; - \; q \;  f(1;t)] \; +
\quad\]
\[+ \; p \; f(\lambda - 1;t) \; [1 \; - \; f(-1;t)] \; + \; q \; 
f(\lambda + 1;t) \; [1 \; - \; f(1;t)],    \quad (73)\]
which holds for all $\lambda$ excluding $\lambda = \pm 1$. 
In the limit $\rho_{-} \to 0$, i.e 
when $f(-1;t) \to 0$,  this equation reduces 
 to Eq.(21). In the continuous-space limit Eq.(73)
attains the form
\[\dot{f}(\lambda;t) \; = \; \frac{1}{2} \; \frac{\partial^2 f(\lambda;t)}{\partial \lambda^{2}} \;
- \; [p \; - \; q \;  - \; p \; f(-1;t) \; + \; q \; f(1;t)] \; 
\frac{\partial f(\lambda;t)}{\partial \lambda} \quad (74)\]
which is exactly Eq.(24) with $\dot{X}(t)$ defined by Eq.(72).

Further on, we find that the correlation function $f(\lambda;t)$ 
at the left-hand adjacent to the PBP site (for $\lambda = 1$) obeys
\[\dot{f}(1;t) \; = \; \frac{1}{2} \; (f(2;t) \; - \; f(1;t)) \; + \; p \; f(0;t) \; (1 \; - \; 
f(-1;t)) \; - \quad\]
\[- \; q \; f(1;t) \; (1 \; - \; f(1;t)) \; - \; 2 \; p \; f(1;t) \; (1 \; - \; f(-1;t)) \; +
\quad\]
\[+ \; 2 \; q \; f(2;t) \; (1 \; - \; f(1;t))    \quad (75) \]
Comparing now Eq.(75) with Eq.(73) we have the following condition
on $f(\lambda;t)$ at $\lambda = 1$:
\[\frac{1}{2} \; (f(0;t) \; - \; f(1;t)) \; = \;  q \; f(2;t) \; (1 \; - \; f(1;t)) \; - 
 \;  p \; f(1;t) \; (1 \; - \; f(-1;t))  \quad (76)\]
Next, from  Eqs.(18), (70) and (71) we can derive
\[\dot{f}(- 1;t) \; = \; \frac{1}{2} \; (f(-2;t) \; - \; f(-1;t)) \; - \quad\]
\[ - \; p \; f(-1;t) \; (1 \; - \; 
f(-1;t)) \; -  \; 2 \; q \; f(-1;t) \; (1 \; - \; f(1;t)) \; +  \quad\]
\[ + \; 2 \; p \; f(-2;t) \; (1 \; - \; f(-1;t)) \; + 
  \; q \; f(0;t) \; (1 \; - \; f(1;t)),    \quad (77) \]
which allows us to deduce the boundary condition on  $f(\lambda;t)$ at 
the point $\lambda = - 1$:
\[\frac{1}{2} \; (f(0;t) \; - \; f(-1;t)) \; = \; - \; q \; f(-1;t) \; (1 \; - \; f(1;t)) \; + \quad\]
\[ + \;  p \; f(-2;t) \; (1 \; - \; f(-1;t))  \quad (78)\]

In the continuous-space limit Eqs.(76) and (78) reduce to
\[\left . \frac{1}{2} \; \frac{\partial f(\lambda;t)}{\partial \lambda}\right |_{\lambda = \pm 1}
\; = \; \dot{X}(t) \; f(\pm 1;t),    \quad (79)\]
which represent two boundary conditions for the continuous-space Eq.(74).
Equations (74) and (79),
with the initial conditions 
\[\left . f(\lambda;t)\right |_{t = 0} \; = \; \rho_{+} \; \text{for} \; \lambda < 0, \quad (80.a)\]
\[\left . f(\lambda;t)\right |_{t = 0} \; = \; \rho_{-} \; \text{for} \; \lambda > 0, \quad (80.b)\]
and the boundary conditions
\[\left . f(\lambda;t)\right |_{\lambda \to \infty} \; = \; \rho_{-}, \quad (81.a)\]
\[\left . f(\lambda;t)\right |_{\lambda \to - \infty} \; = \; \rho_{+}, \quad (81.b)\]
constitute a closed system of equations which allows to  compute $X(t)$
and  the density profiles for arbitrary relation between $p$ and $q$, as well as for arbitrary
$\rho_{+}$ and $\rho_{-}$.

\section{Solution of dynamical equations in the general case $\rho_{-} \geq \rho_{+} \geq 0$.}

In this section we will derive explicit results for the dynamics of the mean displacement
of the PBP and also for the density distribution around it. As it was done in the previous sections,
we will discuss separately the behavior in the case when the HDP expands, compressing
the LDP, and when, on the contrary, the LDP
and the external force $F$
compress the HDP.

\subsection{Expansion of the high-density phase.}

We again
set $X(t) = A \sqrt{t}$ and suppose first that $A \geq 0$. Conditions
 at which such a behavior takes
place will be specified below. 
For $\lambda \geq 1$ (past
the PBP) we then
have
\[f(\omega) \; = \; \frac{\rho_{-}}{1 + I_{+}(A)} \;  \{1 \; + \; A^{2} \; 
\int^{\omega}_{0} dz \; \exp(-\frac{A^{2}}{2}
(z^{2} - 2 z)\}, \quad (82)\]
where $\omega = (\lambda - 1)/ A \sqrt{t}$ and $I_{+}(A)$ is defined in Eq.(10). 
In front of the PBP,
i.e. for $\lambda \leq -1$, 
the scaled density profile is given by
\[f(\theta) \; = \; \frac{\rho_{+}}{1 - I_{-}(A)} \;  \{1 \; - \; A^{2} \; 
\int^{\theta}_{0} dz \; \exp(-\frac{A^{2}}{2}
(z^{2} + 2 z)\}, \quad (83)\]
in which we have denoted 
$\theta = - (\lambda + 1)/ A \sqrt{t}$ and $I_{-}(A)$ is made explicit in Eq.(10). 
The density distributions $f(\omega)$ and $f(\theta)$ for different values of the parameters $A$,
$\rho_{\pm}$ and
$p$ ($q$) 
are depicted in Figs.5,6 and $11$ respectively.

Equations (82) and (83)
contain the parameter $A$, which has not yet been   specified. To determine
$A$ we take advantage of Eq.(72) which yields the following condition on the local
densities at the sites adjacent to the PBP position:
\[q \; (1 \; - \; f(\lambda=1;t)) \; = \; p \; (1 \; - \; f(\lambda= - 1;t))   \quad (84)\]
Upon substitution of Eqs.(82) and (83) into the latter equation we find that $A$ (in case when $A \geq
0$) obeys the following transcendental equation:
\[\frac{q \; \rho_{-}}{1 + I_{+}(A)} \; - \; \frac{p \; \rho_{+}}{1 - I_{-}(A)} \; = \; q \; - \; p,
  \quad (85)\]
which generalizes our Eq.(34) and also the result of \cite{burb} 
(Eq.(4) of the present paper)
for the case when the 
particle densities from the left and from the right of the PBP are different
and the density of the LDP is not zero. 
One directly verifies that Eq.(85) reduces to Eq.(34)
 when we
set $\rho_{+} = 0$, while setting $\rho_{+} = \rho_{-}$ we recover  Eq.(4). 

Let us now find the conditions under which the parameter $A$ is positive, i.e.
 the HDP expands. To do this, we simply 
notice that when $A = 0$ both $I_{+}(A)$ and $I_{-}(A)$ are equal to zero, which means
that the "critical" relation between $p$, $q$ and $\rho_{\pm}$ is:
\[q \; (1 \; - \; \rho_{-}) \; = \; p \; (1 \; - \; \rho_{+}) \quad (86)\]
Equation (86) implies that $A$ vanishes, (i.e. the LDP and the HDP are in equilibrium with each other), when the probability of the PBP to
 go towards the HDP times the density of vacancies in this phase is exactly equal to the probability
of going towards the LDP times the density of vacancies in this phase. 
When $p (1  -  \rho_{+}) \geq q  (1  -  \rho_{-})$ the HDP expands.

We note that Eq.(86) was previously obtained in \cite{fer}
 from the analysis of the stationary states
in a one-dimensional lattice gas placed in a 
finite box of length $L$. 
By explicit calculation of
the distribution function of the PBP
position in the general case 
when the numbers of the lattice gas particles
 from the right and from the left of the
PBP are not equal, it was found \cite{fer}
that in the limit $L \to \infty$ 
the PBP is localized at point $X_{0}$, 
which divides the system in proportion given by Eq.(86).

Equation (86) can  also be rewritten
using the definition of the external force $F$. Upon some algebra, we find then 
that the 
critical force at which both phases are in equilibrium with each other
is given by
\[F_{c} \; = \; \beta^{-1} \; ln(\frac{1 - \rho_{+}}{1 - \rho_{-}}) \quad (87)\]

Now, let us discuss the behavior of the parameter $A$
 in the limit when $A$ is small
or large, and calculate the diffusivity of the PBP 
in the critical case $A = 0$.  
In the limit of small $A$, i.e. when 
$p$, $q$ and $\rho_{\pm}$ are close to their "critical" values  determined
by Eqs.(86) and (87), 
both $I_{\pm}(A) \approx \sqrt{\pi/2} A$. 
Substituting these expressions into  Eq.(85)
we find
\[A \; \approx \; \sqrt{\frac{2}{\pi}} \; \frac{p (1  -  \rho_{+}) 
- q  (1  -  \rho_{-})}{q \rho_{-} + p \rho_{+}}, \quad (88)\]
which is valid when $A \ll 1$. Eq.(88) allows for the computation of the PBP mobility, which we
determine following the arguments presented in  Section IV as 
\[\mu(t) \; = \; lim_{F \to F_{c}} \; \frac{X(t)}{(F - F_{c}) \; t} \; = \quad\]
\[ = \; t^{-1/2} \; 
lim_{F \to F_{c}} \; \frac{A}{(F - F_{c})} \quad (89)\]
Substituting  Eq.(88) into  Eq.(89) and taking the limit $F \to F_{c}$, we find 
\[\mu(t) \; = \; \beta \; \frac{(1 - \rho_{-})(1 - \rho_{+})}{(\rho_{-} + \rho_{+} - 2 \rho_{-}
\rho_{+})} \; \sqrt{\frac{2}{\pi t}}, \quad (90)\]
which yields, by virtue of Eq.(45), the result presented in Eq.(7). Eq.(7) generalizes the classical
result in Eq.(1) for the situation in which the mean particles densities for both sides of the
tracer particle are different from each  other. One can directly verify that Eq.(7) reduces to Eq.(1)
when $\rho_{-} = \rho_{+}$, while setting $\rho_{+} = 0$ we recover our previous result in Eq.(48).
In Fig.12 we compare our analytical prediction in Eq.(90) against the results of Monte Carlo
simulations,  which shows that an approximate approach developed here represents a fair description
of the PBP dynamics.

Next, in Section  IV we have demonstrated that 
the prefactor $A$ diverges when $p \to q$. Consequently, we 
can expect that even in the presence of the LDP the prefactor $A$ can attain large values when
$\rho_{+} \ll 1$
and $p/q \to 1$. Setting in Eq.(85) $q = p$ and using 
the expansion in Eq.(65) we find from Eq.(85) that $A$ is defined in the limit $\rho_{+} \to 0$
by
\[A \; \approx \; \sqrt{2 \; ln(\rho_{-}/\rho_{+})},    \quad (91)\]
i.e. $A$ grows as a square-root of the logarithm of $\rho_{+}$ when $\rho_{+} \to 0$.

Finally, we estimate the behavior of 
the ratio $\delta$ of the particle densities immediately
past and in front of the PBP. At zero moment of time
this ratio is evidently $\delta =  \delta_{0} = \rho_{-}/\rho_{+}$. Our results
in Eqs.(82) and (83) suggest that after some transient period of time the density profiles
past and in front of the PBP attain stationary
 forms with respect to the variable $\omega = (\lambda -1)/X(t)$. Consequently,
 we have that,  as the time evolves,
the ratio of the particle densities immediately past and in front of the PBP
 tends to a constant value
\[\delta \; = \; \delta_{0} \; \frac{1 - I_{-}(A)}{1 + I_{+}(A)} \quad (92)\]
Eq.(92) holds for arbitrary values of $A$. In the asymptotic limits when $A$ is small or large,
we find from Eq.(92) the following explicit asymptotic forms for $\delta$:
\[\delta \; \approx \; \delta_{0} \; (1 \; - \; (\frac{\pi}{2} - 1) \; A^{2} 
\; + \; ... \; ), \; \text{when} \; A \ll 1, \quad
(93)\]
and
\[\delta \; \approx \; \delta_{0} \; \frac{exp(- A^{2}/2)}{\sqrt{2 \pi} A^{3}}
\; \text{when} \; A \gg 1 \quad (94)\]
Therefore, the parameter 
$\delta$, as it could be expected intuitively, is always less than $\delta_{0}$.
Complete dependence of $\delta$ on the parameter $A$ is presented in Fig.14.

\subsection{Compression of the high-density phase.}

Consider now the behavior in the regime when the LDP and the applied
force compress the HDP. Setting $X(t) = - B \sqrt{t}$, where $B$
is supposed to be a positive constant, we find from Eqs.(74) and (79) that the particle
density for $\lambda \geq 0$ (i.e. at sites $X < X(t)$) obeys Eq.(60), in which the
variable $\omega$ is defined as $\omega = (\lambda - 1)/B \sqrt{t}$ and the parameter
$A$ is replaced by $B$. From the other side of the PBP, 
i.e for $\lambda \leq 0$,
we have
\[f(\theta) \; = \; \frac{\rho_{+}}{1 \; + \; I_{+}(B)} \; \{1 \; + \; B^{2} \; \int^{\theta}_{0} dz \; \exp( -
\frac{B^{2}}{2} (z^{2} - 2 z))\}, \quad (95)\]
where the scaled variable $\theta = - (\lambda + 1)/B \sqrt{t}$. The function $f(\theta)$ is depicted in Figs.5
and 6. 
Substituting Eqs.(60) and (95) into  Eq.(84) we arrive at the following transcendental
equation for the parameter B:
\[\frac{q \; \rho_{-}}{1 \; - \; I_{-}(B)} \; - \; \frac{p \; \rho_{+}}{1 \; + \; I_{+}(B)}
\; = \; q \; - \; p     \quad(96)\]
Eq.(96) thus generalizes the result in Eq.(62) for  the case of the non-zero particles density
in the LDP.

Now, noticing that  
 Eqs.(96) and (85) can be cast into one another 
by the substitution 
$\pm I_{\pm}(A) \to \mp I_{\mp}(B)$, 
we  can construct a 
general equation for the parameter 
$\alpha(F)$ in Eq.(3). This equation 
is presented in Eq.(6)
and holds for arbitrary relation 
between  $\rho_{\pm}$ 
and $p/q$, describing hence both the
expansion and the compression regimes. 
In Fig.13 we present the numerical
 solution of Eq.(6), plotting $\alpha(F)$ as a 
function
of the ratio $p/q$ for different 
values of $\rho_{-}$ and $\rho_{+}$.

Finally, we analyze the behavior of the parameter $\delta$, which is defined as
the ratio of the particle density immediately past the PBP and the particle density
immediately in front of the PBP. Using Eqs.(60) and (95) we find
\[\delta \; = \; \delta_{0} \; \frac{1 + I_{+}(B)}{1 - I_{-}(B)} \quad(97)\]
Numerical plot of $\delta(B)$ is presented in Fig.14.

Asymptotic behavior of the parameter $\delta$ in the 
limits when $B$ is small or large readily follows from our
Eqs.(93) and (94).  Here we have
\[\delta \; \approx \; \delta_{0} 
\; (1 \; + \; (\frac{\pi}{2} - 1) 
\; B^{2} \; + \; ... \; ), \; \text{when} \; B \ll 1, \quad
(98)\]
and
\[\delta \; \approx \; \delta_{0} \; \sqrt{2 \pi} \; B^{3} \;  exp(B^{2}/2),  
\; \text{when} \; B \gg 1,   \quad (99)\]
which means that in the compression regime the parameter $\delta$ is always greater
than $\delta_{0}$.

\section{Monte Carlo simulations.}

In order to check our analytical predictions, 
derived in terms of a mean-field approximation, we
have performed
  Monte
Carlo simulations of the process defined in the begining of Section II. 
The simulation algorithm was defined as follows: 

We constructed first a one-dimensional regular lattice
of unit spacing and length $2L+1$, sites of which were labelled by integers
of  the interval $[-L,L]$. In all simulations we took $L = 10^{3}$. 
At the zero moment of the MC
time the particles were placed randomly on the
lattice with the prescribed mean 
densities and the constraint that  two
particles can never simultaneously occupy the same site. 
To do this,  
we have called, for each lattice site from the interval $[-L+1,-1]$ 
independently,
a random number from the interval
$[0,1]$. In case when 
 the  random number produced by the generator 
was less  that $\rho_{-}$ a particle was created on this lattice site.
In case when the random number was greater than $\rho_{-}$ the site was
 left empty. The same routine was performed
 for the sites with positive numbers $[1,L-1]$; 
here a particle was created at the
corresponding site in case when the random number was less than $\rho_{+}$
and the site was left empty if the random number was greater than $\rho_{+}$.
The phase boundary  particle was
placed at the origin. Additionally,  
we have prescribed that the sites $X = \pm L$
are occupied by particles. The particles at theses sites 
$X = \pm L$ are made immobile, blocking the lattice from both
sides and preventing other particle to leave the system.

The subsequent particle 
dynamics employed in our simulations
follows the definitions of Section 2 closely. We call for a random integer 
number $X$
from the interval $[-L+1,L-1]$. Here three different events may take place:

(i) If the site $X$ is vacant, a new site is considered.

(ii) If the site $X$
is occupied by a particle, we first increase the MC time by unity and then
let the particle choose, 
at random,  a potential jump
direction. This is done again by calling a 
random number from the interval $[0,1]$. 
If the random
number is less than $0.5$, the
particle attempts to jump to the site $X -1$; otherwise, 
it attempts to jump to the site $X + 1$.
The jump is  fulfilled if at this moment
of the MC time the adjacent site in the chosen 
direction is vacant (not occupied by any other particle or the PBP). 
Otherwise, the particle
remains at $X$. 

(iii) If the site $X$ appears to be
 occupied by the PBP, we increase the MC time by unity and
consider a random number from the interval $[0,1]$; in case when this number  
is less than the prescribed value $q$, the PBP attempts to jump to the site $X - 1$.
Otherwise, it attempts to jump to the site $X + 1$. The jump is fulfilled if 
at this moment
of MC time the adjacent site in the chosen direction is vacant. Otherwise, the
PBP remains at $X$.

In simulations we have followed the time
 evolution of several different properties:
the PBP displacement, squared displacement (under the critical conditions)
and 
the occupations of the sites $X = \pm L
\mp 100$. Time behavior of these properties
 was plotted versus the
"physical time" $t$, which is the time needed
 for each particle to move once, on
average, or in other words, 
$t$ = MCtime/number of particles. 
We have observed that for all values of the
 parameters $\rho_{\pm}$ and $q$, used in
our simulations, the 
stationary regime in which the ratio $X_{r}(t)/\sqrt{t}$
approaches a constant value 
is established for displacements of order of
$200$ lattice units. To get the spatially resolved behavior, in computation of the PBP displacement
each realization of the process was interrupted 
at the moment when the absolute value 
of the PBP displacement reaches the value of $500$ lattice. 
For calculation of the mean displacement 
we used typically $10^{2}$ realizations 
for each set of parameters $\rho_{\pm}$ and $q$. 
Results of these simulations are presented in Figs.9 and 10.
Further on, computing the squared displacement
we interrupted each realization of the process
at the moment when the span of the PBP trajectory 
is equal to $10^{2}$. Mean-square displacement was obtained
by averaging over $2 \times 10^{3}$ realizations. 
Results for mean-square displacement of the PBP are
presented in Figs.4 and 12. In all cases, we have obtained remarkably good
agreement between our analytical predictions and simulation results. 
Finally, the measurements of the occupations of the sites $X = \pm L
\mp 100$ were performed 
in order to be sure 
that the perturbances created by the PBP
do not spread during the simulation time through the whole system
and do not lead to artificial 
behaviors associated with the finite-size effects.
We have observed that actually the mean densities of these sites
don't vary with time and are equal to the unperturbed values $\rho_{\pm}$.

\section{Conclusions.}

To conclude, we have examined in terms of a mean-field-type  approach
the dynamics of the phase boundary propagation in a one-dimensional
hard-core lattice gas which was initially put into a non-equlibrium, "shock"-like 
 configuration
 and then allowed to
evolve in time by particles attempting to hop to neighboring unoccupied sites. 
The "shock" configuration means that particle mean densities from the left and from the
right of the origin are different.
All particles of the lattice gas, except the particle
separating the high- and the low-density phases, 
have symmetric hopping probabilities, while the phase boundary particle 
is subject to a constant force $F$ and has asymmetric hopping probabilities. 
We have shown that the mean displacement of the PBP follows the generic
law $X(t) = \alpha(F) \sqrt{t}$, 
in which the parameter $\alpha(F)$ can be both positive and negative,
depending on the relation between the magnitude of the 
 force and the initial mean
densities. 
This prefactor is determined implicitly, in a form of the transcendental 
Eq.(6)  for
arbitrary magnitude of the force and arbitrary relation  between the 
particle densities in the high- and low-density phases.  
In several asymptotic limits we find explicit
formulae for the prefactor. Further on, we have shown that when $F$ is equal to the
critical value $F_{c}$, Eq.(87),  the parameter $\alpha(F)$ is exactly equal to zero. In
this case the high- and the low-density phases are in equilibrium with each other. We
have found that here the mean-square displacement $\overline{X_{r}^{2}(t)} $ of the PBP
 follows $\overline{X_{r}^{2}(t)} \sim \gamma \sqrt{t}$, i.e.
shows a sub-diffusive behavior. The form of 
the prefactor $\gamma$ is determined explicitly, Eq.(7).  
Our analytical findings are in a very good
agreement with the results of numerical simulations.

\vspace{0.5cm}

The authors wish to thank J.L.Lebowitz and R.Kotecky for helpful and 
encouraging discussions. Financial support from the FNRS and the COST Project D5/0003/95
is gratefully acknowledged.

\end{document}